\documentstyle[11pt,aaspp4]{article}
\def\et{et al.\ }
\def\xte{{\it RXTE}}
\def\hst{{\it HST}}
\def\asca{{\it ASCA}}
\def\iue{{\it IUE}}
\def\euve{{\it EUVE}}
\def\deg{\hbox{$^\circ$}}

\def\fxv{\sigma_{XS}^2}
\def\ls{\mathrel{\hbox{\rlap{\hbox{\lower4pt\hbox{$\sim$}}}\hbox{$<$}}}}
\def\gs{\mathrel{\hbox{\rlap{\hbox{\lower4pt\hbox{$\sim$}}}\hbox{$>$}}}}

\begin{document}
\title{Intensive \hst, \xte\	and \asca\ Monitoring of NGC~3516: \\
Evidence Against Thermal Reprocessing }

\author{Rick Edelson\altaffilmark{1,2},
	Anuradha Koratkar\altaffilmark{3},
	Kirpal Nandra\altaffilmark{4,5},
	Michael Goad\altaffilmark{2},
	Bradley M.\ Peterson\altaffilmark{6},\\
	Stefan Collier\altaffilmark{6},
	Julian Krolik\altaffilmark{7},
	Matthew Malkan\altaffilmark{1},
	Dan Maoz\altaffilmark{8},
	Paul O'Brien\altaffilmark{2},\\
	J.\ Michael Shull\altaffilmark{9},
	Simon Vaughan\altaffilmark{2},
	Robert Warwick\altaffilmark{2}}

\authoremail{rae@star.le.ac.uk}

\altaffiltext{1}{Astronomy Department; University of California; Los 
Angeles, 
CA 90095-1562; USA}

\altaffiltext{2}{X-ray Astronomy Group; Leicester University; Leicester LE1 
7RH; United Kingdom}

\altaffiltext{3}{Space Telescope Science Institute; 3700 San Martin Dr.; 
Baltimore, MD 20771; USA}

\altaffiltext{4}{NASA/Goddard Space Flight Center; Laboratory for High 
Energy Astrophysics; Code 662; Greenbelt, MD 20771; USA}

\altaffiltext{5}{Universities Space Research Association}

\altaffiltext{6}{Astronomy Department; Ohio State University; 180 W.\
18th Ave.; Columbus, OH 43210-1173; USA}

\altaffiltext{7}{Department of Physics and Astronomy; The Johns Hopkins 
University; Baltimore, MD 20771; USA}

\altaffiltext{8}{School of Physics \& Astronomy and Wise Observatory; 
Tel-Aviv University; Tel-Aviv 69978; Israel}

\altaffiltext{9}{Center for Astrophysics and Space Astronomy; University 
of Colorado; Boulder, CO 80309; USA}

\begin{abstract}

During 1998 April 13--16, the bright, strongly variable Seyfert~1 galaxy 
NGC~3516 was monitored almost continuously with \hst\ for 10.3~hr at 
ultraviolet wavelengths and 2.8~d at optical wavelengths, and 
simultaneous \xte\ and \asca\ monitoring covered the same period.
The X-ray fluxes were strongly variable with the soft (0.5--2~keV) X-rays 
showing stronger variations ($\sim$65\% peak-to-peak) than the hard 
(2--10~keV) X-rays ($\sim$50\% peak-to-peak).
The optical continuum showed much smaller but still highly significant 
variations: a slow $\sim$2.5\% rise followed by a faster $\sim$3.5\% 
decline.
The short ultraviolet observation did not show significant variability.

The soft and hard X-ray light curves were strongly correlated with no 
evidence for a significant interband lag.
Likewise, the optical continuum bands (3590~\AA\ and 5510~\AA) were 
also strongly correlated with no measurable lag to 3$\sigma$ limits of 
$\ls$0.15~d.
However, the optical and X-ray light curves showed very different 
behavior, and no significant correlation or simple relationship could be 
found.
These results appear difficult to reconcile with previous reports of 
correlations between X-ray and optical variations and of measurable lags 
within the optical band for some other Seyfert~1s.

These results also present serious problems for ``reprocessing"
models in which the X-ray source heats a stratified accretion disk 
which then reemits in the optical/ultraviolet:
the synchronous variations within the optical would suggest that the 
emitting region is $\ls$0.3~lt-d across, while the lack of correlation 
between X-ray and optical variations would indicate, in the context of 
this model, that any reprocessing region must be $\gs$1~lt-d in size.
It may be possible to resolve this conflict by invoking anisotropic
emission or special geometry, but the most natural explanation 
appears to be that the bulk of the optical luminosity is 
generated by some other mechanism than reprocessing.

\end{abstract}


\section{ Introduction }

Seyfert~1 galaxies emit substantial luminosity over $\sim$8 decades of 
frequency, from the far-infrared through the hard X-rays.
As this is much too broad to be from a single thermal source, this indicates 
that there must be nonthermal (e.g., synchrotron or inverse-Compton) and/or 
multiple emission components.
Comparison between variations in different wavebands can distinguish between 
these possibilities and, in the latter instance, establish relationships 
between different components.
A well-sampled 1~month campaign of simultaneous {\it International
Ultraviolet Explorer} (\iue) and {\it Rossi X-ray Timing Explorer} (\xte) 
monitoring of the Seyfert~1 NGC~7469 found similar overall amplitudes 
in both bands, 
but the X-rays had much stronger short term variations, and correlated 
interband variability could not be detected (Nandra \et 1998).
Done \et (1990) found that NGC~4051 showed no measurable optical 
variations during a 3~d period in which strong X-ray variations were seen.
Recent contemporaneous \xte/optical observations of NGC~3516 over a 
500~d period (Maoz, Edelson \& Nandra 1999) also failed to detect clear 
interband correlation.
On the other hand, apparent correlations between optical/ultraviolet and 
X-ray variations had been claimed in earlier, less well-sampled 
observations of NGC~4151 (Edelson \et 1996) and NGC~5548 (Clavel \et 1992).
These contradictions indicate that the nature of the relationship between 
X-ray and lower energy variations remains to be clarified.

A similar issue is the relation between variations in hard and soft X-ray 
bands.
In a recent {\it Advanced Satellite for Astronomy and Cosmology} (\asca)
survey, Nandra \et (1997) found some cases in which the 
variability amplitudes at soft X-ray energies were larger than those in the 
hard X-rays (see also Turner \et 1999), but these observations were too 
short to estimate interband lags.
However, in simultaneous {\it Extreme Ultraviolet Explorer} 
(\euve), \asca\ and \xte\ monitoring of NGC~5548 and MCG--6-30-15,
Chiang \et (1999) and Reynolds (1999) respectively reported evidence that 
the variations at softer X-ray energies consistently led those at harder 
X-ray energies.
If confirmed, causality arguments would require rejection of 
``reprocessing" models in which the soft X-rays are ``secondary" 
emission produced by passive reradiation of ``primary" hard X-ray
photons in, e.g., an accretion disk.

Finally, the same arguments apply to variations within the 
optical/ultraviolet as well.
Early studies found no evidence for lags between variations within the 
ultraviolet and optical bands, to a limit of $\ls$1~d in NGC~5548 (Krolik 
\et 1991, Korista \et 1995) and $\ls$0.2~d in NGC~4151 (Edelson \et 1996).
The recent NGC~7469 campaign data suggested evidence for 
progressively longer lags, from 0.3 to 1.8~d, between variations at 1315, 
1825, 4845 and 6962~\AA\ (Wanders \et 1997, Collier \et 1998).
Peterson \et (1998) then reanalyzed the optical/ultraviolet data for 
NGC~4151 (Kaspi \et 1996) and reported similar evidence that the 
shortest wavelength ultraviolet variations preceded those at the longest 
optical wavelengths, albeit at a lower confidence level.
These time lags have been interpreted in models in which the ultraviolet 
is reprocessed in an accretion disk to optical photons (Collier \et 1998).

This paper presents the results of simultaneous X-ray, ultraviolet and 
optical observations of NGC~3516, a bright Seyfert~1 galaxy, designed to 
address these issues by sampling much faster than any previous Seyfert~1 
monitoring campaign.
The observations and data reduction are discussed in the next section.
The variability amplitudes and interband lags were then measured, as
discussed in \S\S~3 and 4.
Surprising evidence was found that the optical showed small variations 
that occurred simultaneously within the optical band (to within 
$\ls$0.15~d) and that were not simply related to the much larger X-ray 
variations.
The theoretical implications of these results are discussed in \S~5, 
followed by a summary of the paper's main results in \S~6.

\section{ Observations and Data Reduction }

Simultaneous observations were made of NGC~3516 with \xte, \asca\ and 
{\it Hubble Space Telescope} (\hst) on 1998 April 13--16.
This bright ($m_{v}=12.5$), strongly variable Seyfert~1 nucleus resides 
in an SB0 galaxy at redshift $z=0.0088$.
NGC~3516 is a northern source ($ \delta = +72 \deg $) that lies near the 
pole of the orbit of \hst, making it the only bright Seyfert 1 galaxy 
over which the ``continuous viewing zone'' (CVZ) passes at some time 
during the year.
(The position of the CVZ changes as the satellite's orbit precesses.)
NGC~3516 was visible without interruption during two periods in 1998.
The \xte\ CVZ also passes over NGC~3516 and in fact this source was in the 
CVZs of {\it both} \hst\ and \xte\ in 1998 February, but STScI was unable to 
schedule the \hst\ observations.
Even so, NGC~3516 was in the \hst\ CVZ during the 1998 April observations
and the \xte\ on-source efficiency was also relatively good.

\subsection{ RXTE Data }

\xte\ observed NGC~3516 from 1998 April 13 08:01:08 to April 16 16:10:44 UT.
The current study is restricted to Proportional Counter Array (PCA), 
{\tt STANDARD-2}, 2--10~keV, Layer 1 data because that is where the PCA is 
best calibrated and most sensitive.
Because Proportional Counter Units (PCUs) 3 and 4 were occasionally turned 
off, only data from the other three PCUs (0, 1 and 2) were used.
Good quality data were accepted on the basis of the following criteria: 
the satellite was outside the South Atlantic Anomaly (SAA), the Earth 
elevation angle was $\ge$10\deg, the offset from the nominal optical 
position was $\le$0\fdg02, and {\tt ELECTRON-0} was $ \le 0.1 $.
The last criterion removes data with high particle background rates in the 
PCUs, and the first censors data with high induced particle count rates, 
both of which are conditions under which the background model is relatively 
unreliable.
After screening, the total amount of good data was 142.9~ks.
Background subtraction was performed using the L7-240 model, but the 
systematic errors are still large ($\sim$0.15~ct~s$^{-1}$).
These dominate the uncertainties on all but the shortest time scales.
See Edelson \& Nandra (1999) for further details of the reduction procedure 
as well as a discussion of the \xte\ PCA background.

Light curves were initially extracted with 16 s time resolution.
The data were rebinned on the orbital sampling period (5760~s), with the 
Earth-occultation gaps as the bin edges.
This has the advantage of improving the signal-to-noise and sampling the 
light curve on the shortest available uninterrupted time scale.
The distribution of the $\gs$200 individual points within each bin was used 
to assign standard errors and mean count rates for each orbit.
Thus, these quantities included statistical errors appropriate to 
variability measurement but not systematic effects such as overall
calibration errors that would be appropriate to spectroscopy.
The resultant light curve is shown in Figure~1.

\begin{figure}[h]\vspace{12cm}
\includegraphics{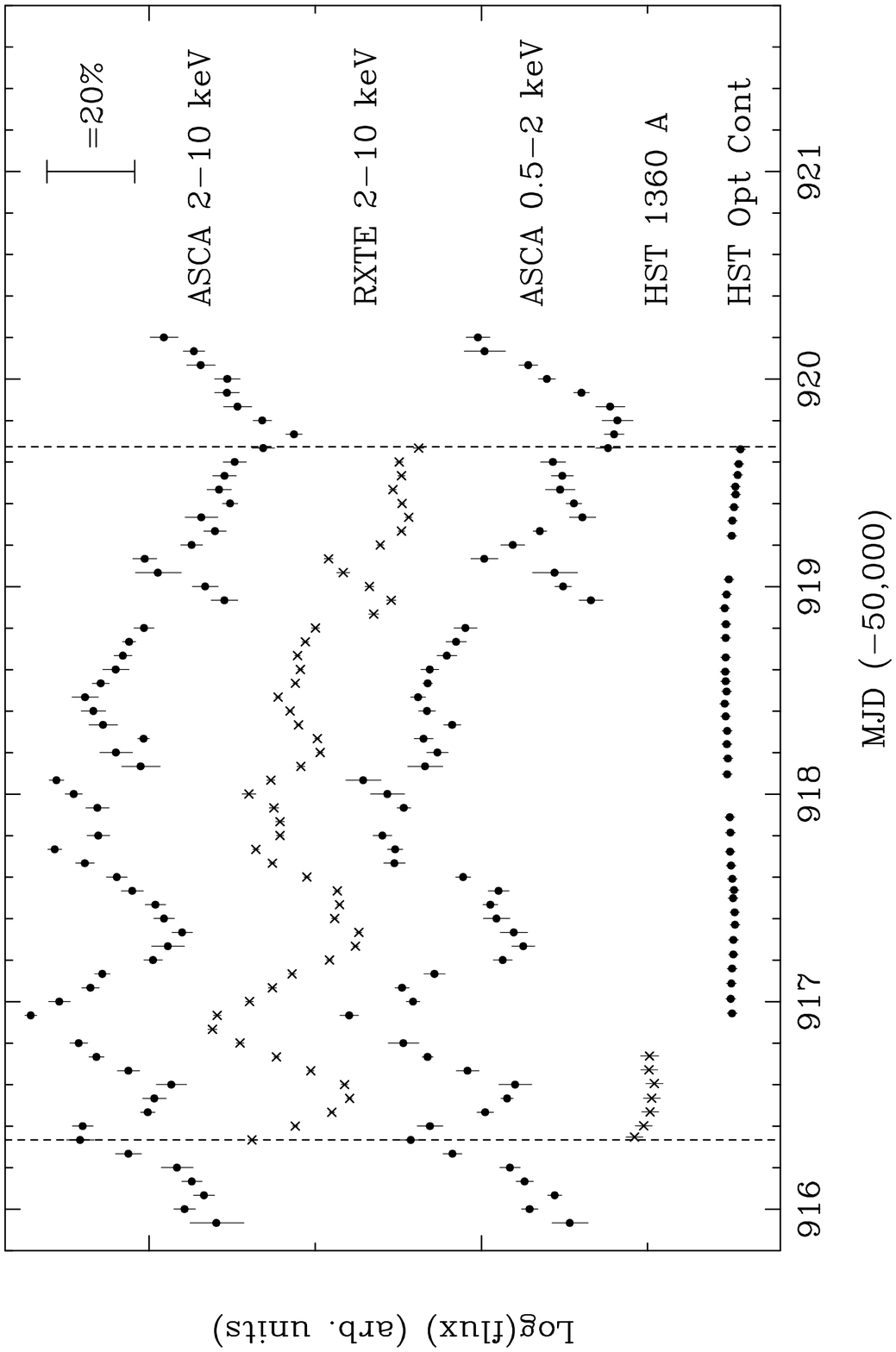}
\vspace{3cm}
\caption{ X-ray, ultraviolet and optical continuum light curves of NGC~3516.
The light curves are, from the top, the \asca\ hard (2--10~keV) band, the 
\xte\ hard (2--10~keV) band, the \asca\ soft (0.5--2~keV) band, the \hst\ 
ultraviolet (1360~\AA) band, and the \hst\ mean optical band. 
The error bar in the upper right shows a factor of 20\% change.
Error bars are 1$\sigma$ statistical uncertainties, except for the
ultraviolet data, which have 1.7\% added in quadrature to account for
the systematic errors.
Note the strong variability in the X-ray bands, but the ultraviolet and 
optical variations are quite weak, as shown in detail in Figure~2.}
\label{fig1}
\end{figure}

\placefigure{fig1}

\subsection{ ASCA Data }

\asca\ observed NGC~3516 from 1998 April 12 22:30:01 to April 17 04:54:01 
UT.
All four detectors were in operation, although here only data from the 
Solid-state Imaging Spectrometers (SIS) are considered.
These were analyzed using standard methods; for further details see Nandra 
\et (1999).
Source counts were extracted from a region centered on NGC~3516 with a 
radius of 4\farcm7.
Data were combined from the two SIS instruments to produce a mean light 
curve, selecting times only when both instruments were deemed to be
collecting data of good quality.
Data were rejected when the angular distance between the pointing position 
and the nominal source position exceeded 0\fdg01, when the satellite was 
passing through the SAA and 16~s thereafter, when the Earth elevation angle 
was less than 10\deg\ (20\deg\ for the bright Earth), when the cut-off 
rigidity was less than 6 GeV/c, when the count rate of the radiation-belt 
monitor (RBM\_CONT) exceeded 500 ct s$^{-1}$, when the pixel number over the 
threshold (Sn\_PIXLn) in the nominal chip exceeded 100, or during the 16~s 
after the satellite passed the day-night terminator.
Hot and flickering pixels were cleaned from the images using standard 
techniques.
Light curves were extracted in 128~s bins in both the hard (2--10~keV) 
and soft (0.5--2~keV) bands.
The \asca\ data were not background subtracted because the flux from
the source filled most of the chip.
We estimate the background contribution to the light curves to be 
no more than a constant $\sim$3\%.
There is no evidence that the SIS background is variable on the time 
scales sampled here, so this made a negligible contribution to the 
overall error and fractional variability level estimates.
These data were rebinned by \asca\ orbit, as described above for the 
\xte\ data.
The resultant light curves are shown in Figure~1.

\subsection{ HST Ultraviolet Data }

\hst\ observed NGC~3516 separately with the Space Telescope Imaging
Spectrometer (STIS) UltraViolet Micro-Anode Multichannel Array
(UV-MAMA) and optical-CCD, covering the wavelength ranges 
1150--1736~\AA\ and 2900--5700~\AA\ respectively.
Because of STIS MAMA limitations (it cannot be used during SAA 
passages), the ultraviolet observations were conducted only during a 
single SAA-free period: 1998 April 13 07:55:14 to 18:17:37 UT.  
The G140L grating (1150--1736~\AA) was used in time-tagged mode with a 
$ 52\arcsec \times 0\farcs5 $ slit for a total exposure time of 32.8 ks.  
To maximize the period in the CVZ, the observations were obtained using the 
smaller bright Earth limb avoidance angle of 16\deg.  
(The increased airglow-induced background is negligible in the UV-MAMA.)  
NGC~3516 was re-acquired at the beginning of each orbit but no 
wavelength calibration (WAVECAL) observations were obtained in order to 
allocate as much time as possible to integration on the target.

The time-tagged data were converted into 6 min integrated images, then
calibrated with CALSTIS v2.0, using the reference files that were
closest in time (and, as it happens, were also the best available).  
The 2-dimensional data were calibrated and extracted into 1-dimensional 
spectra with a 7 pixel extraction (0\farcs35) window.
The relative wavelength calibration accuracy was determined by registering
Galactic C~\textsc{ii} (1334.53~\AA) and Si~\textsc{ii} (1526.71~\AA)
in the orbitally-averaged data.  
There was a maximum motion of 0.3 and 0.2 pixels for the C~\textsc{ii} and
Si~\textsc{ii} lines, respectively, and the 1$\sigma$ variation in the line 
centers was 0.17 pixel, which corresponds to 0.1~\AA.  
The zero point wavelength calibration uncertainty was $\sim$0.7 pixel 
(0.4~\AA).
 
Fluxes were extracted for the 1355--1365~\AA\ continuum band and
C~\textsc{iv} emission line and then light curves were determined by
measuring the mean and standard error on the $\sim$15 individual 6~min
data points in each \hst\ orbit, in an analogous fashion to the X-ray
light curves discussed above.  
The resulting light curves are shown in Figures~1 and 2 and the mean 
spectrum, indicating the continuum bands used in the light curve, is shown 
in Figure~3.

There are a number of possible systematic effects that could affect
the relative flux calibration, including thermal fluctuations,
wavelength drift, spacecraft stability and pointing.  
In particular, the G140L flux measurement is a function of temperature due 
to thermal motion of the target (ISR/STIS 98-27).  
This effect is not corrected for in the pipeline re-calibration and the 
temperature variation seen in these data is 4.05\deg C, which corresponds 
to a systematic change of 1.5\% in flux. 
As discussed in \S~3, this can probably account for a large fraction of the 
apparent variability in this relatively short observation.

\begin{figure}[h]
\epsscale{0.6}
\plotone{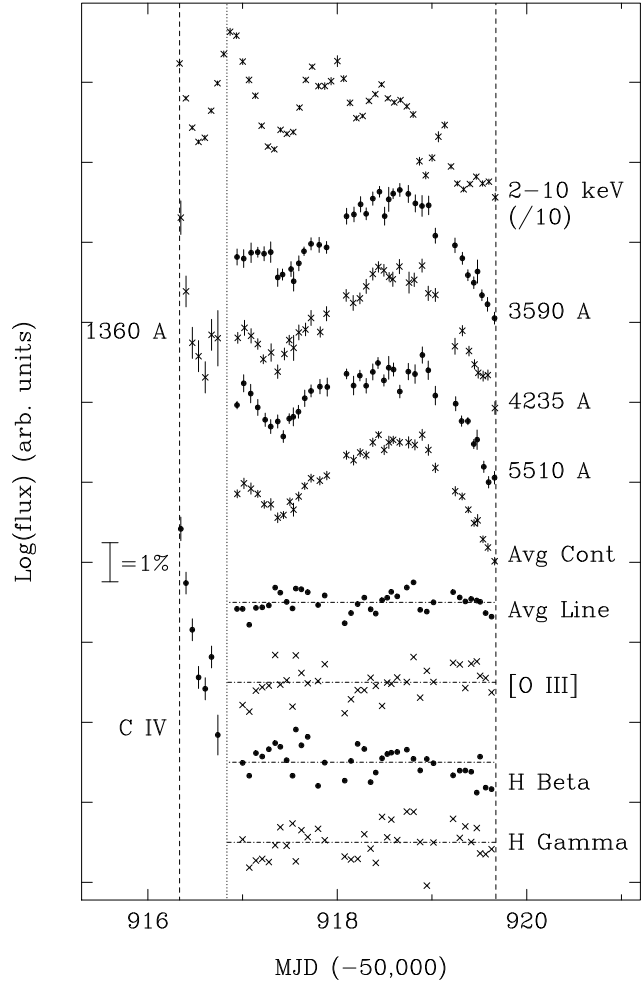}
\caption{Light curves of NGC~3516.
At the top is the \xte\ hard X-ray light curve, scaled down by a factor 
of 10, for comparison with the \hst\ light curves given below.
The ultraviolet light curves (left) are for the 1360~\AA\ continuum (top) 
and C~\textsc{iv} emission line (bottom).
The optical data (right) are, from the top, 3590~\AA, 4235~\AA, 5510~\AA\ 
and average continuum light curves, and the average line, [O~\textsc{iii}], 
H$\beta$ and H$\gamma$ light curves.
The error bars include only statistical errors.
The dashed lines denote the beginning and end of the \xte\ monitoring, 
and the dotted line separates the UV-MAMA from optical CCD measurements.
The horizontal dot-dashed lines show the mean of the emission line light 
curves.
Note specifically that the average optical continuum light curve shows a 
clear $\sim$2.5\% rise followed by a $\sim$3.5\% decline.
During the same period, the average optical emission line light curve is 
flat with an RMS scatter of 0.26\%, indicating that systematic effects 
are not likely to be a problem above this level.}
\label{fig2}
\end{figure}

\placefigure{fig2}

\begin{figure}[h]
\vspace{12cm}
\includegraphics{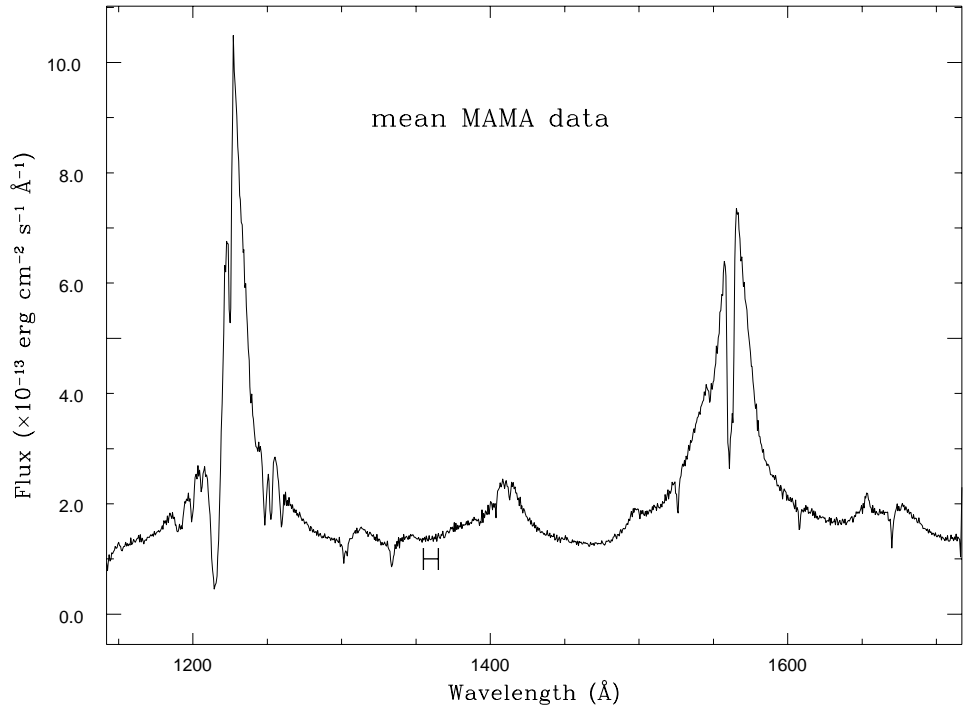}
\vspace{4cm}
\caption{Ultraviolet spectrum of NGC~3516.
The position of the 1355--1365~\AA\ continuum extraction band is 
noted by the small horizontal error bar.}
\label{fig3}
\end{figure}

\placefigure{fig3}

\subsection{ HST Optical Data }

Optical spectra were obtained every 3 min using the STIS CCD/G430L
grating and the $ 52\arcsec \times 0\farcs5 $ slit from 1998 April 13
21:58:11 to April 16 16:59:09~UT.  
Four orbits were lost due to spacecraft/instrument problems, yielding a 
total optical exposure time of 138.2~ks.  
As with the UV-MAMA observations, all observations were obtained using the 
smaller bright Earth limb avoidance angle of 16\deg, and no WAVECALs were 
taken in order to increase the on-target integration time.  
The increased airglow-induced background can be effectively corrected for 
in the CCD/G430L grating.
 
As with the STIS MAMA data, the STIS CCD data were calibrated and
extracted into 1-dimensional spectra with a 7 pixel extraction
(0\farcs35) window.
The 1-dimensional spectra were measured as both 3~min and orbital averages.  
The 3~min data were used to measure continuum fluxes and errors in the same 
fashion as with the previous data sets, but this was not done for the 
line fluxes because wavelength drift would have required manually adjusting 
the wavelength scales for almost 1,000 spectra.  
Instead, wavelength adjustment was performed for the 38 orbitally-averaged 
spectra using Galactic lines.
Initially, the $1 \sigma$ variation was 0.17 pixels (0.46~\AA), and
the maximum motion was 0.6 pixels (1.6~\AA).  
After wavelength correction the maximum motion in the lines was 0.031 
pixels (0.08~\AA).  
These data were used to measure the line fluxes, but because the 
orbitally-averaged data were used, errors were not estimated.

STIS CCD observations were performed in ALONG-SLIT mode. 
This involved trailing the source parallel to the slit (that is, in the 
spatial direction) to enable clean removal of bad pixels and 
cosmic-ray events.  
Superposed on this motion of the source along the slit were additional 
small ($\pm$0.5 pixel) motions perpendicular to the slit axis (that is, 
in the dispersion direction), as a consequence of dithering of the 
parallel Wide Field/Planetary Camera (WFPC2) observations. 
These were performed to reduce the effects of small-scale non-uniformity 
in the WFPC2 detector, increasing the dynamic range and effective spatial 
resolution of the WFPC2 observations. 
We have investigated the effects of this dithering and find that there
is no systematic relation between the position and either continuum or line
fluxes, so the dithering is not a likely source of systematic error.

In the calibrated spectra, fluxes were measured for the
3575--3600~\AA, 4223--4245~\AA, 5500--5525~\AA\ continuum bands and
[O~\textsc{iii}] (5137--5255~\AA), H$\beta$ (4861--4946~\AA) and 
H$\gamma$ (4350--4430~\AA) emission lines.
In addition, mean optical continuum fluxes (defined as the harmonic 
mean of the fluxes in the three continuum bands) and mean line fluxes 
(the harmonic mean of the three line fluxes) were also determined for 
each orbit. 
These light curves are presented in Figures~1 and 2 and the mean 
spectrum is presented in Figure~4.  

Narrow band images of NGC~3516 in [O~\textsc{iii}] (4959 and 5007~\AA) 
show a biconical extended (5\arcsec) emission-line region with position 
angle of 25\deg\ (Golev \et 1995).
Due to roll angle constraints, the STIS aperture was aligned at
position angle 98\deg.
The slit position is almost perpendicular to the extended narrow 
emission-line region.
This means that in principle, small variations could be induced in the
derived narrow emission-line flux due to variations in spacecraft 
roll angle ($\sim 2$\deg\ maximum), although the [O~\textsc{iii}] light
curve shows no evidence of excess scatter (e.g., above that seen in
the Balmer lines), so this is apparently not a problem.

\section{ Variability Amplitudes }

To quantify the amplitude of variations in different bands, we computed the 
fractional excess variance, $\fxv$, defined as
\begin{equation}
\fxv = { S^2 - \langle \sigma_{err}^2 \rangle \over \langle X \rangle^2 },
\end{equation}
where $ \langle X \rangle $ is the mean flux, $ \langle \sigma_{err}^2 
\rangle $ is the mean square error, and $S^2$ is the measured variance of 
the light curve (see Nandra \et 1997).
It is intended as a measure of the intrinsic source variability power 
during a given time interval, corrected for the effects of measurement 
noise and normalized to the mean flux.
No error has been estimated for the emission line measurements, so the
quoted values of $\fxv$ are in these cases upper limits.

Table~1 summarizes these results for the X-ray, ultraviolet and optical 
data.
It shows that the X-rays were strongly variable, and that the fractional 
excess variance measured with both \xte\ and \asca\ in the hard X-ray
band (2--10~keV) is smaller than that measured in the softer X-ray band
(0.5--2~keV) with \asca.
The larger soft band variability is also apparent in Figure~1.

\placetable{tab1}

The optical/ultraviolet continuum variations were much weaker than those 
in the X-rays.
In the optical, the excess variance in the continuum is approximately twice 
that measured in the [O~\textsc{iii}] and H$\beta$ lines and 50\% larger 
than that in the H$\gamma$ line.
Furthermore, the apparent variations in the lines are not coherent, but 
those in the continuum are, with all three continuum bands showing a slow 
rise of $\sim$2.5\% followed by a faster decline of $\sim$3.5\%.
This behavior is not seen in any of the emission line light curves.
The key independent test is provided by the mean emission line light curve.
It is completely flat, with an RMS standard deviation of 0.26\%.
We conclude that the systematic errors are no larger than this. 
This in turn indicates that these small optical continuum variations
are real.
While these would be improbably small errors for ground-based
observations, we note that previous observations have demonstrated that
\hst\ is capable of such high-precision monitoring (e.g., Welsh \et
1998).

In the ultraviolet, $\fxv$ is somewhat larger for the C~\textsc{iv} line 
than for the 1360~\AA\ continuum.
It is unlikely that C~\textsc{iv} would vary by this large an amount 
($\sim$5\% peak-to-peak) in 10~hr; furthermore, the line and continuum light 
curves have very similar shapes.
Also, as discussed in \S~2.3, the ultraviolet data suffer from instabilities 
not seen in other bands (in particular, detector gain changes induced by 
thermal variations).
This all suggests that the ultraviolet continuum variability could very 
likely have resulted from systematic effects not included in the measured 
(statistical) errors.
Thus, we are forced to conclude that these data do not give
unambiguous evidence of significant ultraviolet variations.

\section{ Interband Correlations }

In order to further examine the relation between variations in different 
bands, temporal cross-correlation functions were measured using both the 
interpolated correlation function (ICF; White \& Peterson 1994) and the 
discrete correlation function (DCF; Edelson \& Krolik 1989).
Errors on the ICF lags were estimated using the bootstrap method of 
Peterson \et (1998).

\subsection{ Correlations Within the X-Rays }

Within the hard X-ray band (2--10~keV), the \xte\ and \asca\ light curves 
were highly correlated ($ r = +0.97 $) with no hint of any lag.
A zero-lag correlation diagram for these data is shown in Figure~5.
The smaller error bars and higher count rates indicate that in this band,  
\xte\ is the superior satellite for monitoring bright Seyfert~1s 
like NGC~3516.
The very good correlation indicates that systematic errors in the 
background model to not significantly affect either data set.

\placefigure{fig5}

Figure~6 shows the temporal cross-correlation functions between the two 
X-ray bands (top), within the optical (middle) and between the optical 
and X-rays (bottom).
The \xte\ hard -- \asca\ soft ICF reaches a maximum correlation
coefficient of $ r_{max} = +0.95 $ for a lag of $ \tau = -0.02 $~d.
The two highest points in the DCF are at zero lag and --1 orbit 
(--0.067~d), and a smoothed parabolic fit is centered on 
$ \tau = -0.020 {+0.018 \atop -0.022} $~d (in the sense that the \xte\ 
hard X-ray variations would lead the \asca\ soft X-ray variations).
These values are less than half of a single orbital bin and we consider 
this result to be consistent with the null hypothesis, that is, that 
there is no measurable interband lag with a 3$\sigma$ limit of 
$ \tau \ls 0.07 $ ~d.

\placefigure{fig6}

\subsection{ Correlations Within the Optical }

No significant lag was seen within the optical band either.
Along the longest wavelength baseline, 3590 -- 5510~\AA, the DCF peaked 
at zero lag and the ICF centroid was  
$ \tau = -0.012 {+0.053 \atop -0.052} $~d (see Figure~6).
Likewise, over the 3590 -- 4235~\AA\ baseline, the DCF peaked at zero lag 
and the ICF centroid was at $ \tau = -0.002 {+0.035 \atop -0.047} $~d 
(not shown).
In both cases, the correlation is strong ($ r_{max} = 0.906 $ and 0.944, 
respectively).
The 3$\sigma$ upper limits on any possible lags are $ \tau \ls 0.15 $~d.

\subsection{ Correlations Between X-ray and Optical Variations }

The peak in ICF for the hard X-ray (\xte\ 2--10~keV) and mean optical 
continuum light curves is at $ \tau = -0.21 { +0.07 \atop -0.34 } $~d, 
but the maximum correlation coefficient is only $ r_{max} = +0.53 $ 
(see Figure~6).
This would be significant at the $ P \approx 0.0008 $ level if it was 
the result of only one trial, and if the data were all independent.
Because there were 38 trials, the significance is lower and $P$ must 
be multiplied by 38, yielding $ P \approx 3 $\%.
Furthermore, the red noise character of the fluctuation Power Density 
Spectra (PDS) means that the measurements are not independent.
This, in turn means that the (already marginal) significance must be 
further reduced (see Maoz \et 1999).
Monte-Carlo simulations indicate that the corrected probability is 
$ P \approx 24 $\%, which is not significant.
In fact, the most significant value of $r$ is the {\it anticorrelation} 
with $ r_{min} \ls -0.8 $ for $ \tau = +0.6 $ to +1.3~d.
This anticorrelation is certainly not predicted by any model, and the
fact that it is so broad further suggests that it is a spurious effect
as discussed above.
Thus, we must conclude that these data contain no clear evidence for 
correlated X-ray/optical variability.

To sum up these results, the optical and X-ray variations show no 
significant interband correlation, while within each of the bands, the 
intraband variations are highly correlated with no measurable delays longer 
than the strong upper limits of a few hours.
These findings have important physical implications, as discussed below.

\section{ Discussion }

\subsection{ Reprocessing Models }

Much attention has been given to reprocessing models for the optical
emission from Seyfert~1 galaxies in which an X-ray continuum source
irradiates relatively dense and cool material which, in turn, emits 
thermal radiation at longer (optical/ultraviolet) wavelengths (e.g., 
Guilbert \& Rees 1988, Rokaki \et 1992). 
There is a good body of X-ray spectral evidence in support of
this model: 
many Seyfert galaxies show evidence for a strong ``Compton bump" in the
X-ray spectrum, a signature of reflection from an absorbing medium
(Lightman \& White 1988, Pounds \et 1990, George \& Fabian 1991).
Any absorbed X-ray flux must also be re-emitted. 
A likely candidate for the putative absorber (but not the only one, 
e.g., Krolik \et 1994, Ghisellini \et 1994) is thermally-emitting 
matter close to the central black hole (e.g., an accretion disk). 
Any such thermalizing source should radiate in the optical/ultraviolet.
There is also a strong theoretical prejudice that the primary energy
release should occur just outside the marginally stable orbit around a
black hole as the result of flow through an accretion disk.
It is plausible that both the X-ray and optical/ultraviolet source could
be located in that vicinity.
Moreover, specific Comptonization models making use of this geometry
result in rough agreement with observed spectra (e.g., Haardt \& Maraschi
1991).

This model (and any other involving a central driver, e.g., Krolik \et
1991, Clavel \& Courvoisier 1991) has important implications for the
variations seen in the optical continuum.
On time scales longer than the light-crossing time of the reprocessing
region, the optical/ultraviolet variations should follow those in the
(driving) X-ray band, but at shorter time scales, they should be
smoothed out by light travel-time effects.
Because most of the short-wavelength optical/ultraviolet continuum arises 
at smaller radii than most of the longer-wavelength continuum, this model 
also implies a time delay between the variations at short and long 
wavelengths.

The details of both the ``smearing" of rapid fluctuations and the 
interband delays depend on details of the system's structure.
However, their general character may be illustrated by a simple model: 
Suppose that most of the flux at wavelength $\lambda$ is thermal
radiation from material with temperature $T \sim hc/(k\lambda)$, 
(e.g., as in a conventional accretion disk) where $h$, $c$ and $k$ 
are physical constants.
Then the area of this region is $ A \sim \lambda L_\lambda/(\sigma
T^4) $, where $ L_\lambda$ is the luminosity at that wavelength and
$\sigma$ is a physical constant, and the associated length scale is 
$ R \propto (\lambda L_\lambda)^{1/2} \lambda^2$.
If the signal propagation speed is $v_s$ ($= c$ in a reprocessing model),
both the most rapid fluctuation that can be reproduced at wavelength
$\lambda$ and the delay at that wavelength are of order
\begin{equation}
\tau (\lambda) \sim  R/v_s \sim \sigma^{-1/2} \left({k \over hc}\right)^2
\left(\lambda^5 L_\lambda\right)^{1/2}.
\end{equation}
For an externally irradiated accretion disk, the luminosity emitted per 
unit area is proportional to $R^{-3}$ and the temperature distribution is 
$L_\lambda \propto \lambda^{-7/3}$, so that $\tau \propto \lambda^{4/3}$.
Moreover, if the central source luminosity is proportional to the 
accretion rate, then $\tau \propto R \propto (M \dot M)^{1/3}$ 
(Shakura \& Sunyaev 1973).
Combining these, the overall relation is 
$\tau \propto (M \dot M)^{1/3} \lambda^{4/3} $ (Collier \et 1998,
Peterson \et 1998).

\subsection{ NGC~3516 and NGC~7469 }

Time delays of this sort have in fact been reported in the Seyfert 1 galaxy 
NGC~7469 (Wanders \et 1997, Collier \et 1998).
Other multiwavelength monitoring programs have generally not been 
sufficiently well sampled or of sufficient duration (or both) to expect 
detection of these delays, although there is marginal evidence for such a 
lag between the shortest-wavelength ultraviolet and the longest-wavelength 
optical continuum bands in NGC~4151 (Peterson \et 1998).
The lags are roughly consistent with the predicted relation
$ \tau \propto (M \dot M)^{1/3} \lambda^{4/3} $.
However, even in NGC~7469, the magnitudes of relative continuum lags 
$\tau = 0.36^{+0.11}_{-0.17}$~d between 1315~\AA\ and 1825~\AA, and 
$\tau = 1.7^{+1.1}_{-0.8}$~d between 1315~\AA\ and 6962~\AA\
(Kriss \et 1999), are uncomfortably close to the mean sampling of 
the light curves (0.17~d in the ultraviolet and 1.0~d in the optical).

Fitting the $\tau \propto \lambda^{4/3}$ relation to the NGC~7469 data
(as done by Collier \et 1998) yields a predicted lag between 3590~\AA\ 
and 5510~\AA\ of $\tau = 0.55 \pm 0.27 $~d.  
If the central mass can be estimated from the formula 
$ M \propto v^2 \tau_e $ (e.g., Peterson \& Wandel 1999), where $v$ is 
the broad line velocity width and $\tau_e$ the emission-line lag,
then the ratio of virial masses NGC~7469 and NGC~3516 is 
$ M_{3516}/M_{7469} = 0.29 $.
Furthermore, if the accretion rate scales as the luminosity, 
$ L_{3516}/L_{7469} = \dot{M}_{3516}/\dot{M}_{7469} = 0.50 $.
This crude scaling predicts the relative time delay between the 
3590~\AA\ and 5510~\AA\ variations of NGC~3516 will be $ \tau = 0.28 $~d.
Although this way of estimating $M \dot M$ is very uncertain,
relatively large uncertainties can be tolerated because the dependence 
of $\tau$ on $ M \dot{M} $ is so weak.

Alternatively, one could assume that the bolometric luminosity has a 
fixed ratio to the ultraviolet luminosity, and that the Eddington ratio 
$L/L_E$ is the same in NGC~3516 as in NGC~7469.
In this case, $ \tau \propto L^{2/3} $, which predicts that the relative 
time delay between the 3590~\AA\ and 5510~\AA\ variations of NGC~3516 
will be $ \tau = 0.31 $~d.

Hence, the observed upper bound of $\tau \ls 0.15 $~d in NGC~3516 appear 
inconsistent with the predicted value.  
However, it is unclear how much stress should be laid upon this 
conflict, given the numerous model dependencies built into the
derivation and the fact that the NGC~7469 result is thus far unique 
and less than definitive.
it should be noted that a similar discrepancy would have been found upon 
comparison with almost any conventional accretion disk model.

\subsection{ Implications for X-ray/Optical Correlations }

The combination of a lack of measurable correlation between the X-ray
and optical light curves and the synchronicity within the optical band
presents even more serious general problems for the reprocessing model
discussed above, independent of any possible disagreements between the
scaling from the NGC~7469 results.
The problem is that each of these results implies a limit on the size
of the reprocessing region that is incompatible with the other.

The lack of response to the X-rays indicates that the light-crossing
time of the optical/ultraviolet reprocessing region is of order or
larger than the duration of the simultaneous X-ray/optical monitoring.
Otherwise, the light curves should show some correlation if the
optical/ultraviolet is in fact driven by the X-rays.
Since this experiment ran 2.8~d, this would indicate that the
reprocessing region would have to be $\gs$1~lt-d in size.
In fact, much longer term monitoring also failed to show the X-rays 
leading the optical variations, on delay time scales of weeks to months 
(Maoz \et 1999).
This in fact indicates that the reprocessing region should be
light-weeks across or larger.

On the other hand, the observation of significant optical variability
with no lags between the bands down to time scales of $\ls$0.15~d
yields an upper limit on the size of the reprocessing region.
For the most straightforward geometry, this upper limit in the lag
yields an upper limit on the radius of the emitting region.
Assuming the relation derived earlier for the stratified temperature
structure of an externally-irradiated $\alpha$-disk (Shakura \& 
Sunyaev 1973), $ T \propto R^{-3/4} $.
By Wein's law, $ T \propto 1/\lambda_{max} $, so $ 
R \propto \lambda_{max}^{4/3} $.
For a ratio of peak wavelengths 3590~\AA/5510~\AA, the ratio of the
distances of the rings emitting most strongly at 3590~\AA\ and
5510~\AA\ is 1.9, that is, the radius of the 5510~\AA-emitting ring is
1.9 times that of the 3590~\AA-emitting ring.
An upper limit of $\ls$0.15~lt-d for the distance between them 
corresponds to an upper limit of $\ls$0.3~lt-d for the radius of the
entire system.
Furthermore, if the optical continuum is produced in the same region 
as the iron K$\alpha$ line, the relativistic effects observed in the 
line profile of NGC~3516 (Nandra \et 1999) would also argue for an 
origin much closer to the central source.
Thus, quite independent of the conflict with an extrapolation of any
NGC~7469 result, these data create difficulties for the simplest
reprocessing models.
 
There are a number of ways to get around this contradiction.
One possible explanation is that the flux of X-rays striking the disk 
is too small to significantly affect its output.
The observed monochromatic luminosity at 3590~\AA\ is about twice that 
at 2~keV, but since the X-ray variability amplitude is $\sim$20 times 
that in the optical, the variable monochromatic power in the X-rays is 
$\sim$10 times that in the optical.
However, the scaling of the monochromatic to integrated luminosities 
of the X-ray and optical/ultraviolet components are 
not well-determined, and this may decrease this ratio.
Even if the X-ray luminosity is great enough to affect the optical, 
it might be that most of the X-rays are actually directed away from
the disk, e.g., if the solid angle subtended by the disk seen by the
X-ray source is small (Dove \et 1997) or if it is moving away from 
it at relativistic speeds (Beloborodov 1999).  
However, all of these models contradict the assumption that the observed 
optical variations are driven by the variable X-ray source.

Another possibility is that the reprocessor is smooth, but that the X-ray 
source is patchy, flaring and of non-negligible size compared to the 
reprocessor (Stern \et 1995). 
There is some evidence for this, especially the scale-free character
of the short term X-ray variability (e.g., McHardy \& Czerny 1987),
which argues against a single, coherent source.
In this case, because the geometry is so complex, the temporal
relation between variations cannot be easily predicted.

However, it may be that the most natural explanation is simply that
the optical and X-ray-emitting regions are powered primarily by
different processes.
In this case, there is no reason to assume that a central source
drives the stratification of the putative disk, so individual regions
can change brightness independently, and the low-amplitude optical 
fluctuations could be due to a modulation in the luminosity of regions 
that do not dominate the total flux in these bands.  
For example, if only a single region varied, one would expect 
progressively larger fractional amplitudes of variation toward shorter 
wavelength (as is, in fact, generally observed; Edelson, Krolik 
\& Pike 1990), and these variations would be simultaneous, as observed.

Finally, these results are consistent with the claims of no clear
ultraviolet/X-ray correlation in NGC~7469 (Nandra \et 1998) and 
longer-term optical/X-ray monitoring of NGC~3516 (Maoz \et 1999).
The earlier claims of significant correlations between X-ray and 
optical/ultraviolet variations in NGC~5548 (Clavel \et 1992) and NGC~4151 
(Edelson \et 1996) may have been due to the ``red-noise" character of the 
variations (see, e.g., Edelson \& Nandra 1999), which can lead to an 
overestimate of the significance of a measured correlation (Maoz \et 1999).

\subsection{ Hard/Soft X-ray Variability }

The overall appearance of the variations in the soft and hard X-ray 
bands was similar, with no significant measurable lag between bands.
However, the soft X-rays were a factor of $\sim$30\% more strongly variable 
than the hard X-rays.

As noted in the introduction, there have recently been claimed 
measurements of phase differences (lags) between variations in hard and
soft X-rays, with the soft X-rays leading the hard X-rays by 
$\sim$0.06~d in NGC~5548 (Chiang \et 1999) and by $\sim$0.001~d in 
MCG--6-30-15 (Reynolds 1999, Lee \et 1999).
These are $\sim$0.9 and 0.02 times the fundamental orbital sampling rate.
No error estimates were reported, although subsequent analysis
yielded an error estimate of $ {+0.03 \atop -0.05} $~d on the NGC~5548 
lag (Nowak, priv.\ comm.).

In the current work, we estimated that the lag was not significantly
different from zero, as the hard (2--10 keV) X-rays were seen to lead
the softer (0.5--2~keV) X-rays by $ 0.020 {+0.018 \atop -0.022} $~d.  
That is, the most likely lag seen in NGC~3516 is in the {\it opposite}
sense as that reported for NGC~5548 and MCG--6-30-15.
If real, this would seem to suggest that not all Seyfert galaxies 
exhibit the same type of temporal relationship between hard and soft 
X-ray variations.  
It is not clear what physical process could produce such behavior.

\section{ Conclusions }

This paper reports the results of the most intensive multiwavelength 
Seyfert~1 monitoring campaigns ever undertaken:
continuous, simultaneous monitoring of the Seyfert~1 galaxy NGC~3516 
once every $\sim$96~min, at optical, soft X-ray and hard X-ray wavelengths, 
obtained with \hst, \asca\ and \xte\ respectively, over a 2.8~d period.
The \hst\ data were repeatable at the $\ls$0.26\% level or better.
The observational results were:

\begin{enumerate}

\item 
The X-ray variations were very strong, $\sim$65\% peak-to-peak in the 
soft X-ray band (0.5--2~keV) and $\sim$50\% in the hard X-ray band 
(2--10~keV).
These light curves were highly correlated ($ r = 0.95 $), with no 
measurable interband lag to 3$\sigma$ limits of $ \tau \ls 0.07 $~d.

\item 
The optical continuum bands showed small but significant variations:
a slow $\sim$2.5\% rise followed by a faster $\sim$3.5\% decline.
The variations were highly correlated across the optical continuum 
bands ($ r \ge 0.9 $), with no measurable interband lag (to a 3$\sigma$ 
limit of $ \tau \ls 0.15 $~d between 3590~\AA\ and 5510~\AA).

\item 
Temporal cross-correlation functions gave no evidence for a simple 
relation between the X-ray and optical variations.
The most significant value was the anticorrelation of 
$ r \ls -0.8 $ for $ \tau = -0.8 $ to --1.3~d and the maximum positive 
correlation of $ r = +0.53 $ at $ \tau = -0.21 $~d, which was not deemed 
significant after accounting for interdependence of the data.

\item 
The optical emission lines showed no evidence for variability: a light
curve constructed by averaging [O~\textsc{iii}], H$\beta$ and H$\gamma$ 
line fluxes was flat with 0.26\% RMS dispersion.
Likewise, $\sim$5\% variations seen during the short preceding 
$\sim$10~hr ultraviolet observation were not deemed significant as they 
are most likely due to gain drifts in the MAMA detector or other 
systematic effects.

\end{enumerate}

Earlier monitoring observations reported evidence for wavelength-dependent 
lags in the optical/ultraviolet variations in NGC~7469, in the sense that 
shorter-wavelength variations led those at longer wavelengths by 
$\sim$0.36--1.8~d, with longer lags over longer wavelength baselines.
These were explained by differences in light travel times from the
central source (that illuminates the disk and drives all the
variations) to the hot, inner, ultraviolet-emitting regions and
cooler, outer, optical-emitting regions.
However, this ``reprocessing" model is called into question because 
it predicts optical interband lags that should have been detected in 
these observations of NGC~3516, although an definitive statement 
cannot be made because of the uncertainty in the scaling between 
NGC~7469 and NGC~3516.

The combination of the lack of X-ray/optical correlation and of lags 
between optical bands also presents more serious and general 
difficulties for this reprocessing model, as the sizes indicated by the 
first point are much larger than the upper limits implied by the second.
There are a number of possible model fixes, including anisotropic emission
or source geometry, localized flares, or long processing time scales in
the disk.
However, perhaps the most natural explanation is that the 
unproven assumption of this reprocessing model, that the X-ray
emission powers that at optical/ultraviolet wavelengths, is in error.
Clearly, much theoretical work needs to be done to make the 
reprocessing model fit with this new, emerging picture of Seyfert~1 
interband variability.

More observational work is also required: this was an unexpected result in 
an experiment designed for different purposes.
Now that it has been established that \hst\ is capable of extremely high 
precision relative photometry (see also Welsh \et 1998), it is 
straightforward to design an experiment expressly to search for very 
small lags between optical and ultraviolet variations.
This can eliminate the inherent uncertainties in comparing model 
predictions to the data and resolve the ambiguity concerning the 
relationship between the X-ray through optical variations.

\acknowledgments
The authors appreciate the fine work of the \xte\ project, particularly Evan 
Smith and Tess Jaffe, in scheduling and helping in the reduction of these 
observations.
The authors would also like to thank Andrew Fruchter of STScI for help
in scheduling and determining the times of the dithers. 
This study received financial support from NASA \xte\ grant NAG 5-7315 and 
\hst\ grant GO-07355.
KN acknowledges support from the Universities Space Research Association.
BP and SC acknowledge support from NASA LTSA grant NAG~5-8397.

\clearpage

\begin{figure}[h]
\vspace{12cm}
\includegraphics{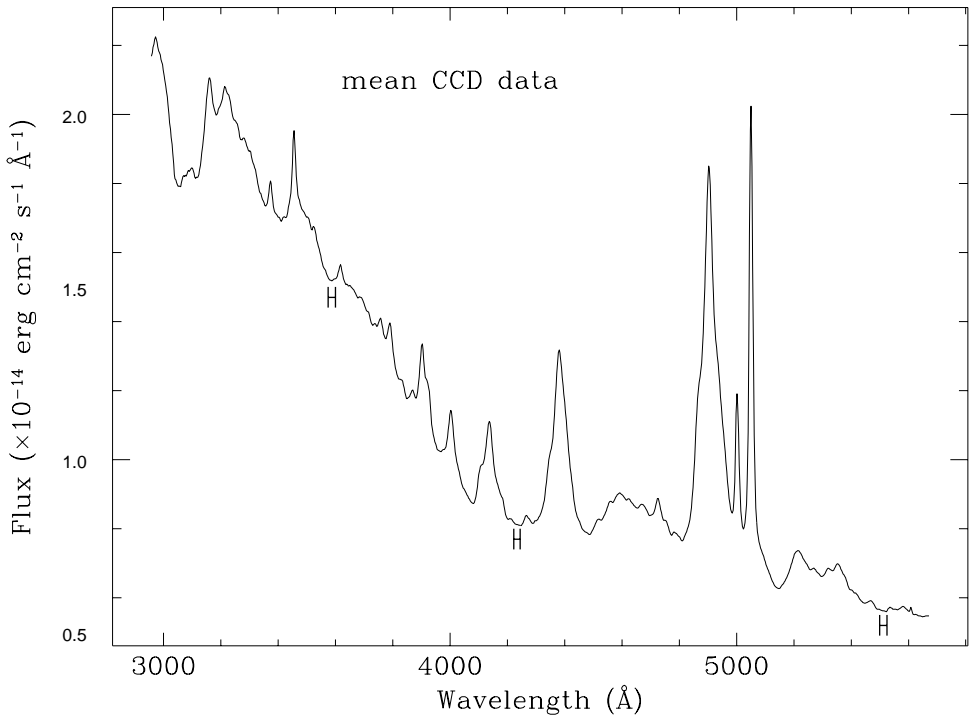}
\vspace{4cm}
\caption{Optical spectrum of NGC~3516.
Positions of continuum extraction bands (3575--3600~\AA, 4223--4245~\AA, 
5500--5525~\AA) are noted by the small horizontal error bars.}
\label{fig4}
\end{figure}

\begin{figure}
\epsscale{1.0}
\vspace{-4 cm}
\plotone{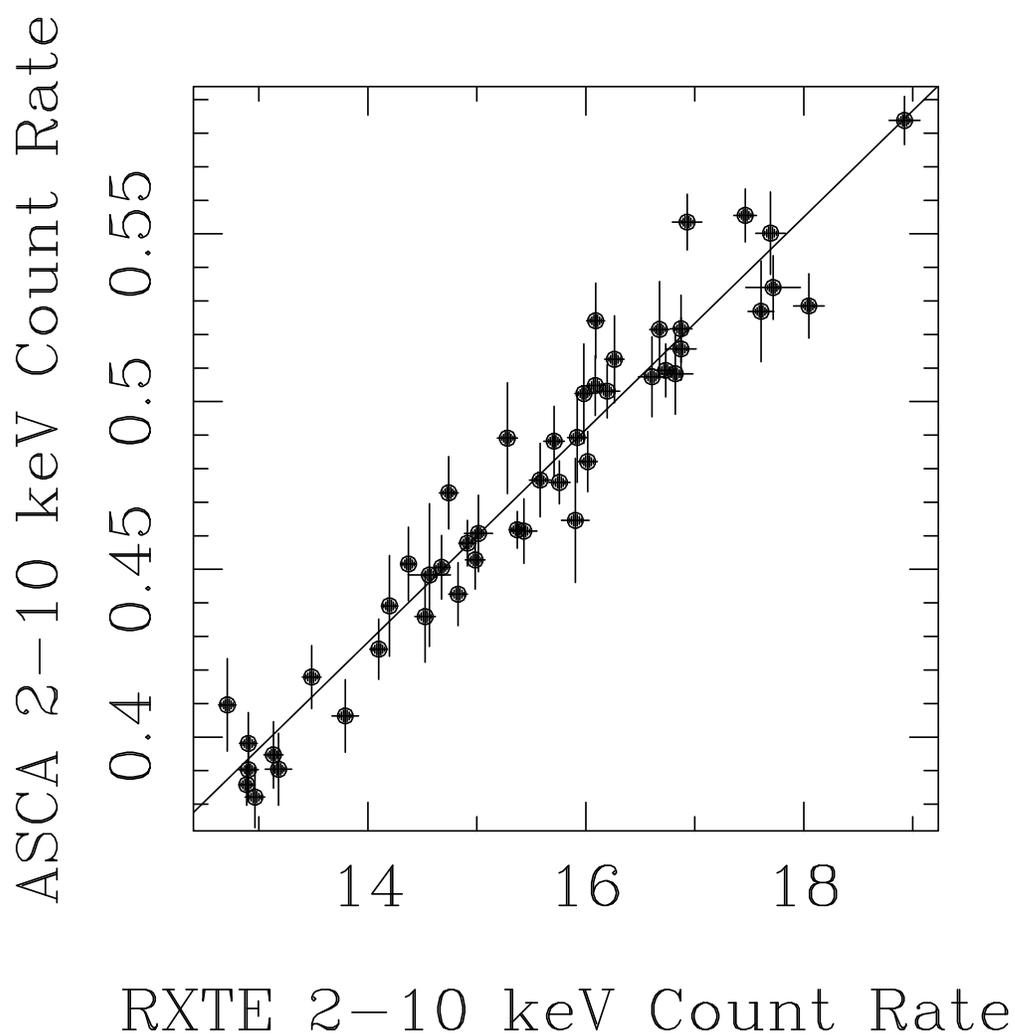}
\vspace{-2 cm}
\caption{Zero-lag correlation plot of 2--10 keV count rates measured from 
\xte\ and \asca.
The solid line is a linear fit to the data, which goes within 0.5$\sigma$
of the origin (not shown).
Note that the \xte\ data have much higher count rates, and much better 
signal-to-noise, than the \asca\ data.}
\label{fig5}
\end{figure}

\begin{figure}[h]
\epsscale{0.7}
\plotone{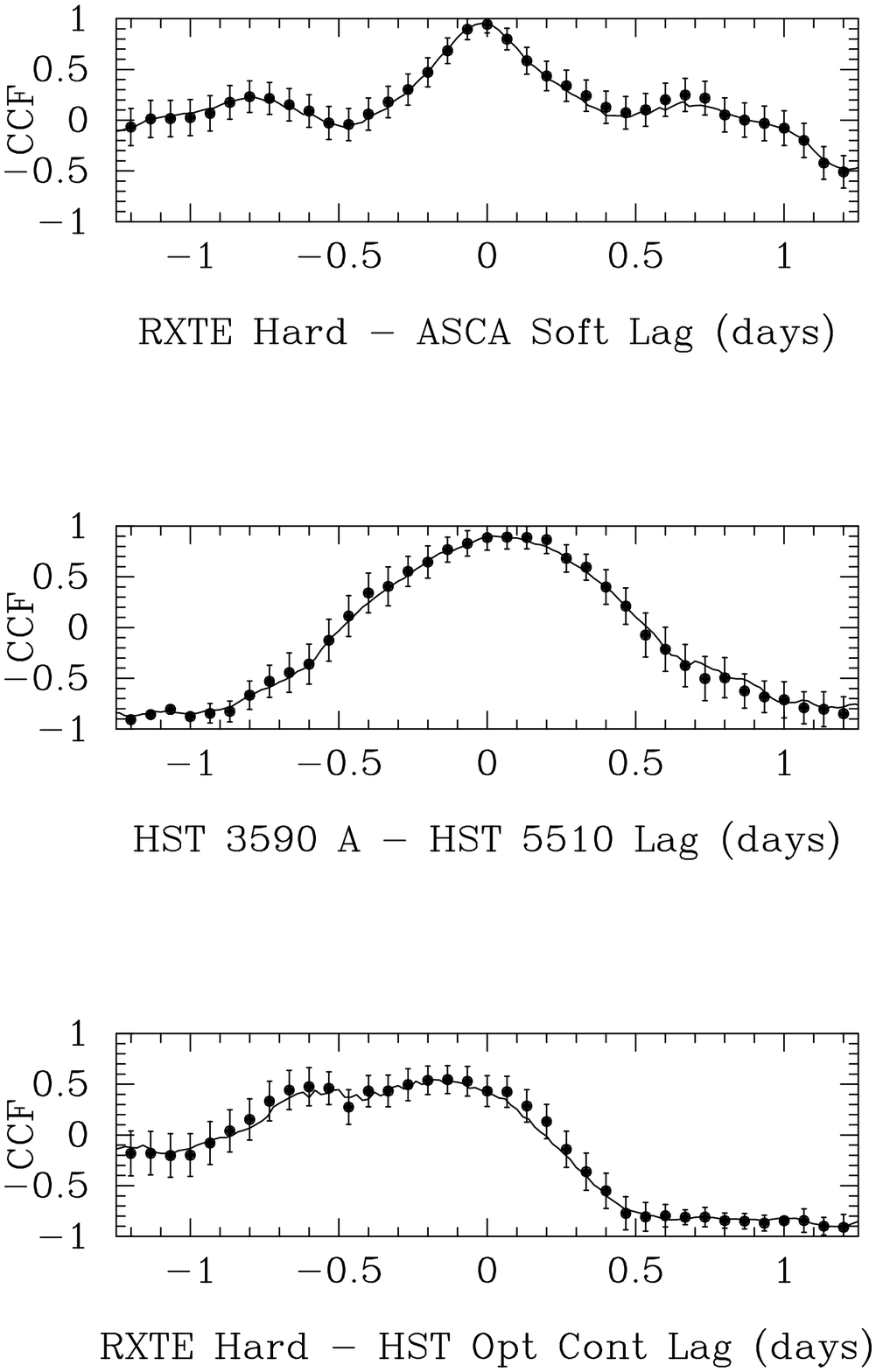}  
\caption{Interband temporal cross-correlation functions.
The solid line refers to the interpolated cross-correlation function, while 
the error bars refer to the discrete cross-correlation function.
At the top is the \xte\ hard -- \asca\ soft correlation function, next is 
the \hst\ 3590~\AA\ -- 5510~\AA\ correlation function, and at the bottom is 
the \xte\ hard -- mean optical continuum correlation function.
In all plots a negative lag means that the first listed band leads the 
second.}
\label{fig6}
\end{figure}

\clearpage

\begin{deluxetable}{ccccccc}
\tablewidth{0pc}\tablenum{1}
\tablecaption{Fractional Variability Levels 
\label{tab1}}
\small\tablehead{
\colhead{Satellite/ } & \colhead{Duration} & \colhead{Number} & 
\colhead{Mean} & \colhead{Fractional} & \colhead{Fractional} & \colhead{} \\
\colhead{Band} & \colhead{(day)} & \colhead{of Orbits} & \colhead{Flux} & 
\colhead{RMS (\%)} & \colhead{Error (\%)} & \colhead{$\fxv$} }
\startdata 
RXTE 2-10 keV          & 3.4 & 51 & 15$^1$   & 10.81 & 0.71 & 0.0114   \\
ASCA 2-10 keV          & 3.4 & 48 & 0.47$^1$ & 11.29 & 2.28 & 0.0119  \\
ASCA 0.5-2 keV         & 3.4 & 48 & 0.51$^1$ & 13.45 & 2.21 & 0.0172  \\
\hline 
HST UV 1360 \AA\ Cont  & 0.4 &  7 & 19$^2$   &  1.39 & 0.45 & 0.000143 \\ 
HST UV C IV Line       & 0.4 &  7 & 30$^3$   &  1.80 &   & $<$0.000277 \\
\hline 
HST Opt 3590 \AA\ Cont & 2.8 & 38 & 15$^2$   &  0.86 & 0.18 & 0.000069 \\
HST Opt 4235 \AA\ Cont & 2.8 & 38 & 8.1$^2$  &  0.94 & 0.17 & 0.000083 \\
HST Opt 5510 \AA\ Cont & 2.8 & 38 & 5.6$^2$  &  0.76 & 0.18 & 0.000052 \\
HST Opt Avg Cont       & 2.8 & 38 & 8.9$^2$  &  0.82 & 0.11 & 0.000066 \\
HST Opt [O~\textsc{iii}] Line   
                       & 2.8 & 38 & 3.7$^3$  &  0.40 &   & $<$0.000016 \\
HST Opt H$\beta$ Line  & 2.8 & 38 & 6.4$^3$  &  0.40 &   & $<$0.000015 \\
HST Opt H$\gamma$ Line & 2.8 & 38 & 2.7$^3$  &  0.44 &   & $<$0.000019 \\
HST Opt Avg Line       & 2.8 & 38 & 4.0$^3$  &  0.26 &   & $<$0.000007 \\
\enddata
\tablecomments{
$^1$X-ray fluxes (count rates) in ct~s$^{-1}$ \\
$^2$\hst\ continuum fluxes in $ 10^{-15}$ erg cm$^{-2}$ s$^{-1}$ \AA$^{-1}$ \\
$^3$\hst\ line fluxes in $ 10^{-12}$ erg cm$^{-2}$ s$^{-1}$ }
\end{deluxetable}

\clearpage

\end{document}